\documentclass[numbers]{elsarticle}
\usepackage{latexsym}
\usepackage{float}
\usepackage[latin2]{inputenc}
\usepackage{graphicx}
\usepackage{ae}
\usepackage{aecompl}
\usepackage{setspace}
\usepackage{booktabs}
\usepackage{longtable}
\usepackage{mciteplus}
\usepackage[hidelinks]{hyperref}

\makeatletter
\def\ps@pprintTitle{%
	\let\@oddhead\@empty
	\let\@evenhead\@empty
	\def\@oddfoot{\centerline{\thepage}}%
	\let\@evenfoot\@oddfoot}
\renewcommand\@biblabel[1]{(#1)} 
\makeatother

\title{The Length and the Width of the Human Brain Circuit Connections are Strongly Correlated}
		
		\author[p]{Dániel Hegedűs}
		\ead{hegedus@pitgroup.org}
		\author[p,u]{Vince Grolmusz\corref{cor1}}
		\ead{grolmusz@pitgroup.org}
		\cortext[cor1]{Corresponding author}
		\address[p]{PIT Bioinformatics Group, Eötvös University, H-1117 Budapest, Hungary}
		\address[u]{Uratim Ltd., H-1118 Budapest, Hungary}

\begin{document}

\begin{abstract}
The correlations of several fundamental properties of human brain connections are investigated in a consensus connectome, constructed from 1064 braingraphs, each on 1015 vertices, corresponding to 1015 anatomical brain areas. The properties examined include the edge length, the fiber number, or edge width, meaning the number of discovered axon bundles forming the edge and the occurrence number of the edge, meaning the number of individual braingraphs where the edge exists. By using our previously published robust braingraphs at \url{https://braingraph.org}, we have prepared a single consensus graph from the data and compared the statistical similarity of the edge occurrence numbers, edge lengths, and fiber counts of the edges. We have found a strong positive Spearman correlation between the edge occurrence numbers and the fiber count numbers, showing that statistically, the most frequent cerebral connections have the largest widths, i.e., the fiber number. We have found a negative Spearman correlation between the fiber lengths and fiber counts, showing that, typically, the shortest edges are the widest or strongest by their fiber counts. We have also found a negative Spearman correlation between the occurrence numbers and the edge lengths: it shows that typically, the long edges are infrequent, and the frequent edges are short.
\end{abstract}

\date{}
	
\maketitle	
	
\section*{Introduction} 

In the twentieth century only very sparse information was available on the anatomical connections between the distinct areas of the human brain. Light- and electron-microscopy staining methods, electrography and in some cases electroencephalography (EEG) were the only available techniques for finding those connections. 

Because of this lack of connection-discovering techniques, anatomical brain morphology studies mostly measured and compared the volumes and the thickness of distinct cerebral regions (i.e., volumetric studies, e.g., \cite{Szalkai2016b}), and the study of the human brain connections could not be developed. 

The study of the anatomical brain circuitry and connections was accelerated dramatically by the application of diffusion magnetic resonance imaging (diffusion MRI) techniques in brain science \cite{Hagmann2008}. Publicly available data sources \cite{LaMontagne2019,McNab2013} and data processing tools (e.g., the Connectome Mapper Tool Kit, CMTK, \cite{Daducci2012}) made it possible to map the network of the connections of the human brain on a macroscopic scale: a new discipline, the connectomics has appeared on the stage of neuroscience.

Our research group has contributed numerous novel techniques for analyzing these braingraphs or connectomes with strict mathematical methods on typically several hundred nodes (or, in other words, vertices). We have computed more than one thousand braingraphs, each in five different resolutions, which are publicly available at \url{https://braingraph.org} \cite{Kerepesi2016b, Varga2020}; developed a new data augmenting method for artificial intelligence applications and used this method for computing 126,000 human braingraphs, each in five resolutions, 630,000 connectomes in total \cite{Keresztes2020}; mapped the individual variability of the circuit connections in \cite{Kerepesi2015a,Szalkai2015a,Szalkai2016} and created a public server at \url{https://connectome.pitgroup.org} visualizing the variabilities of those connections. The ``Budapest Reference Connectome'' computational tool led us to the discovery of the Consensus Connectome Dynamics \cite{Kerepesi2016,Szalkai2016e,Szalkai2016d} and the new method of directing the edges of the connectome \cite{Kerepesi2015b,Szalkai2016d} (the diffusion MRI is not capable of assigning directions to the axonal fibers). We have found that women have significantly better graph-theoretical parameters in their braingraphs than men in \cite{Szalkai2015,Szalkai2016a}, which is independent of the statistical difference of the brain volume of the sexes \cite{Szalkai2015c}. We have mapped the frequently appearing subgraphs in the human connectome in general in the work \cite{Fellner2017}, and, particularly, the frequently appearing {\it complete} subgraphs in \cite{Fellner2019}. 

The human hippocampus is the most frequently studied cerebral area; we have mapped its frequent neighbor sets in \cite{Fellner2018} and discovered its neighbor sets related to high and low human intelligence test scores in \cite{Fellner2019a}. 

The properties of single connections or edges (in contrast to the more complex graph theoretical properties) in the braingraph are rarely examined. 

One of the first studies of the single connections was the work of \cite{Keresztes2022a}, where we considered the fiber numbers in the connections of the braingraph and found dozens of edges whose fiber numbers are characteristic of the sex of the subject: that is, if the fiber number is larger than an edge-specific threshold in one of those connections, then the subject is male, otherwise female, or {\it vice versa}, with good accuracy. It is interesting that those edges, which with higher fiber numbers, imply the male sex (the male implicator edges), are located mostly in the anterior lobes, while the female-implying ones (the female implicator edges) are in the posterior areas. Additionally, the inter-hemispheric implicator edges are mostly male, while the intra-hemispheric implicator edges are mostly female. 

In work \cite{Hegedus2022}, a consensus braingraph was prepared by averaging the edge labels of 1064 subjects.  Then, the averaged labels were ordered by seven features, describing anticipated ``vertex importance'' measures. From the labels of the edges features of the nodes were computed by applying the characteristics of the incident graph edges, describing cerebral connections. It was assumed that an ``important'' vertex was connected to other vertices with ``stronger'' or more ``emphasized'' edges. More exactly, the vertex features examined in \cite{Hegedus2022} were the following ones:

\begin{itemize}
\item the degree of the node, i.e., the number of the connected edges to the node;

\item the sum of the fiber numbers of the edges, which connect to the node;

\item the maximum number of fibers in the incident edges;

\item the average number of fibers in the incident edges;

\item the sum of the fiber lengths in the incident edges;

\item the maximum fiber length in the incident edges;

\item the average fiber length in the incident edges.
\end{itemize}

It was demonstrated in work \cite{Hegedus2022} that all the seven importance measures above yield statistically similar orders of importance among the vertices of the braingraph. 

In the present work, we study a related question as in \cite{Hegedus2022}, but for the edges instead of vertices. Here, we intend to compare the orders of importance on edges instead of the vertices. We will study and compare the edge lengths (described by average fiber lengths), edge widths or strengths (measured by fiber numbers), and edge frequencies or occurrence numbers.

We need to add two notes at this point:

\noindent {\bf Remark 1:} The function of our brain is fundamentally determined by the complex system of connections between neurons. Mankind is unable to study the neuronal level connections of the whole human brain with more than 3 billion neurons (the frontiers of our techniques allow us to study the connections of the fruit fly with 100 thousand neurons in the near future), but we are able to study the connections of the human brain in a much coarser level by magnetic resonance imaging. Here we concentrate on the characteristics of the macroscopic, MRI-discovered  single edges of the brain and not on the more involved graph-theoretic properties, involving several edges, like in studies \cite{Kerepesi2016,Szalkai2016e,Szalkai2016d,Fellner2018,Fellner2019a,Kerepesi2015a,Szalkai2015a,Szalkai2016}. 

\noindent {\bf Remark 2:} In the work of \cite{Hegedus2022}, we examined not the single edges but the ``bundles'' of edges, which were incident to the same node. The characteristics of those edges were inherited by the nodes, and this way, we have shown that the seven orders of ``node importance'' listed above are all statistically similar. Here, we make comparisons between the orders on the edges and not on the nodes.  In our consensus graph, we have 1015 vertices and 99,171 edges, and we compare the orders on those 99,171 edges. Therefore, now we will have larger deviations and not-so-strong correlations as in the case of the orders on 1015 nodes in \cite{Hegedus2022}, but even in these much larger ordered sets, we will find statistically strong correlations in the braingraph.

\section*{Methods}

\subsection*{Braingraph construction}

In the present work, we use the data of the \url{https://braingraph.org} resource, where our research group has published thousands of human braingraphs \cite{Kerepesi2016b, Szalkai2016d, Varga2020}. The present work applied the ``1015 nodes set, 1064 brains, 1 000 000 streamlines, 10x repeated \& averaged'' set for constructing the consensus graph ``One single averaged consensus connectome with 1015 nodes'', freely available at the site \url{https://braingraph.org/cms/download-pit-group-connectomes/}. While the construction of the graphs is described in detail in \cite{Varga2020} and \cite{Hegedus2022}, we summarize here the construction steps succinctly for completeness:

\begin{itemize}
\item[(i)] The data source for building the graphs is the diffusion MRI repository of the 1200 Subjects Data Release of the Human Connectome Project (HCP) \cite{McNab2013}. For the graph constructions, the ``re-preprocessed'' 3T diffusion data were used. 

\item[(ii)] The Connectome Mapper Toolkit \cite{Daducci2012} was applied for graph computation, with 10-times repetitions and averaging of the probabilistic tractography step, with the Lausanne2008 atlas. The nodes of the resulting graphs correspond to the anatomical regions of interest, and two nodes are connected by an edge if the tractography step identified at least one streamline between the regions. If more than one streamline is identified then we connect the two regions by one edge, but the edge labels will depend on the averaged attributes of the streamlines.

\item[(iii)] An edge connecting the vertices $A$ and $B$ carries three labels: the number of fibers or streamlines connecting $A$ and $B$ (also called edge thickness), the average length of the streamlines and the average fractional anisotropy of the streamlines. 

\item[(iv)] Steps (i)-(iii) were applied for the diffusion MRI data of 1064 human brains, each with 1015 anatomically identified nodes, available at  \url{https://braingraph.org/cms/download-pit-group-connectomes/}. 

\item[(v)] From the 1064 braingraphs, one single consensus graph was computed as follows: The nodes were the 1015 vertices of the graphs, and each edge in the consensus graph carried the labels of average fiber number (or thickness) and the average edge length of the 1064 braingraphs. 

\end{itemize}

We have applied two averaging methods: the ordinary average computing of the edge labels and the non-zero average computing. For a given edge $e$ in subject $i$, let $\ell^e_i$ denote the edge label, and let $n=1064$ be the number of braingraphs in the consensus graph. Then the ordinary average of the labels of $e$ is defined as $${1\over {n}}\sum_{i=1}^n \ell^e_i,\eqno{(1)}$$ while the non-zero average is $${1\over {n_e}}\sum_{i=1}^n \ell^e_i,\eqno{(2)}$$
where ${n_e}$ is the number of the non-zero $\ell^e_i$ labels for $i=1,2,\ldots,n$. Since the
labels were not all-zero for any edge, which appear in at least one subject, the division by ${n_e}$ is valid.

For the edge lengths, only the non-zero edge lengths were averaged (in order to compute only the average length of the {\it existing} connections; otherwise, the average length values were misleading because of the non-edges in some subjects); i.e., we have applied formula (2).

For the fiber numbers, we have computed the ordinary average values by formula (1), since it is a notable event if between two vertices  there is no connection in some of the subjects.

The resulting consensus graph is freely available at \url{https://braingraph.org/cms/download-pit-group-connectomes/}. The graph contains 1015 vertices, each labeled with the name of an anatomical gray matter area, and 99,171 edges corresponding to their connections.

\subsection*{Edge orderings by fiber number, length, and frequency}

The main focus of the present contribution is the comparison of the edge orderings of the consensus graph according to their three different weights: the fiber number, the edge length, and the frequency or the occurrence number. In point (v) above, we described the computation of the fiber number and the edge length labels. The frequency (or the occurrence number) is simply the number of appearances of the same edge in the 1064 graphs of the subjects: its value is between 0 (meaning no occurrence) and 1064 (meaning that the edge appears in all the braingraphs). 

We have ordered all the 99,171 edges according to these three labels in decreasing order. When the orderings are by the occurrence number, then many of these 99,171 edges carry the 1, 2, 3 and 4 labels as occurrences (we have 99,171 edges and 1064 possible occurrences), and by visualizing those orderings, scattered or gapped figures would be the result. Therefore, we have chosen the edges which are present in at least 5 subjects, and visualized only the relative orderings of those edges. This way, there are no artifacts on the figures. With this restriction, we have finally processed 67,281 edges. 
 
Next, we compared the pairs from these orderings graphically (cf. Figures 1, 2, and 3) and statistically using Spearman's rank coefficient. In the Figures only those 67,281 edges are visualized which are present in at least 5 subjects.

\subsection*{Statistical analysis: the Spearman's rank correlation coefficient}

The Spearman correlation coefficient \cite{Spearman} is used for quantifying the correlation of two random variables. The more frequently used Pearson correlation is adequate for describing linear relations between random variables, while the Spearman correlation can also quantify more general non-linear relations by describing their ``joint monotonicity'': it compares two orderings of the values of two random variables and the positions in the orderings of the corresponding values of the two random variables are quantified  \cite{Hegedus2022}. The Spearman coefficient is usually denoted by $\varrho$. Similarly to the Pearson correlation, $|\varrho|\leq 1$ and  $\varrho$ value around 1 means strong positive correlation, $\varrho$ value around -1 means strong negative correlation, while $\varrho$ around zero means weak or no correlation. 

As we remarked in \cite{Hegedus2022}, for large $n$ values, the coefficients with smaller absolute values can also be significant. Therefore, the $p$-values accompanying the Spearman correlations also need to be observed.  We note that we write p value $0.0$ (to be distinguished from $0$) below if the computed p value is less than $10^{-300}$.

\section*{Results and Discussion}

A very informative visualization of the relative positions of the same item in two different orderings is given in a diagram, where on axis $x$ the first decreasing order $(x_1,x_2,\ldots,x_n)$, on axis y the second decreasing order $(y_1,y_2,\ldots,y_n)$ is listed, and a point with coordinates $(x_i,y_j)$ visualizes an item whose index is $i$ in the first and $j$ in the second order. 

In our case, the items are the edges of the consensus braingraph, $n=67,281$, and the three orderings are defined as decreasing orders of fiber numbers, edge lengths, and the number of occurrences.

 Figures 1,2 and 3 compare these orderings.

\subsection*{Occurrence -- fiber number comparison} 

Our Figure 1 depicts the relation between the order of edge occurrence numbers and fiber numbers.

\begin{figure}[H]
			\includegraphics[width=12.5cm]{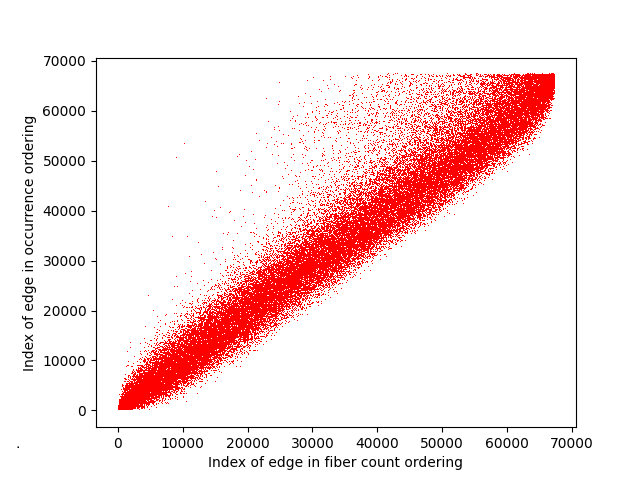}
\caption{The comparison of edge indices in the orderings (in decreasing order)  of occurrence number and fiber number in the consensus graph. In the diagram, we have demonstrated the relative positions of 67,281 connections, each of which are present in at least 5 subjects, in the two orderings of the consensus braingraph. Every point corresponds to an edge (67,281 in total), which has an $x$-coordinate equal to its index in the decreasing fiber number ordering, and the $y$-coordinate equals to its index in the frequency-ordering. 
On the figure there is a very strong Spearman correlation between the edge occurrence number and the fiber count number (or thickness) of the edges: the Spearman coefficient is $0.97$, and the $p$-value is $0.0$. In other words, the thickest edges are present in most braingraphs. 
It is important to notice that there are only a very few frequent and thin (i.e., non-thick) edges in the graph, since most of the lower right corner is empty.
}
\end{figure}

\subsection*{Fiber length -- fiber number comparison} 

Figure 2 shows the relative orders according to the fiber counts and fiber lengths.

\begin{figure}[H]
\includegraphics[width=12.5cm]{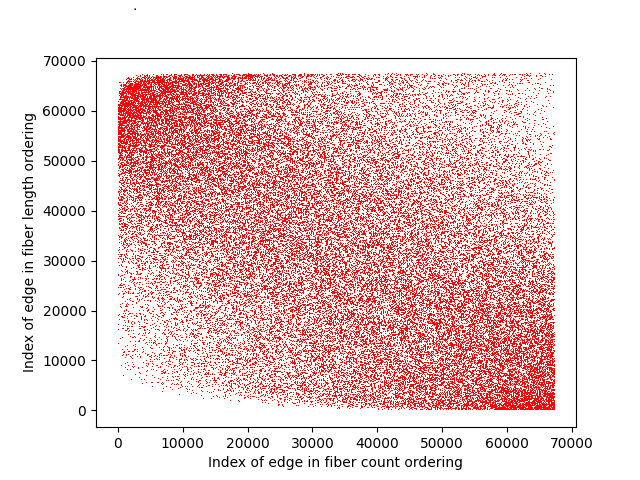}
\caption{The comparison of edge indices in the orderings of length and fiber number. The fiber lengths were averaged by formula (2), while the fiber numbers were averaged by formula (1). 
In the diagram, every point corresponds to an edge, which has $x$-coordinate equal to its index in the thickness-ordering, and $y$-coordinate equal to its index in the length-ordering. 
Clearly, a negative Spearman correlation can be observed between the orders: that is, the long edges are typically thin, and the short edges are typically thick, but the correlation is not so strong as in Figure 1. 
In the figure the lower left corner is empty; that is, there are no thick and very long edges. The Spearman coefficient is $-0.51$, the $p$-value is $0.0$.}
\end{figure}

\subsection*{Occurrence -- fiber length comparison} 

\begin{figure}[H]
\center
\includegraphics[width=12.5cm]{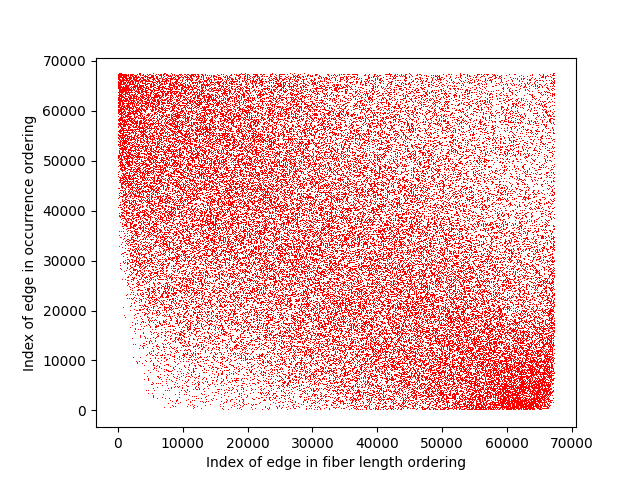}
\caption{The comparison of edge indices in the orderings (in decreasing order)  of occurrence number and fiber length in the consensus graph. In the diagram, we have corresponded each of the 67,281 connections to a point whose x coordinate is determined by its position in the order of the edge-length, and the y coordinate to its position in the order of the occurrences. 
A negative correlation is observable on the chart (the Spearman coefficient is $-0.47$, the p-value is $0.0$), meaning that typically frequent edges are usually short ones, and long edges are typically not frequent. The virtually empty lower left corner indicates that very few long and frequent edges exist.
}
\end{figure}

\subsection*{Control experiment}

As a control experiment, we have taken the consensus graph, removed the measured edge labels, and substituted them by randomly generated labels for all edges, comprising the fiber number, the edge length and the occurrence labels. Next, we generated the edge orderings according to the random labels, and computed the Spearman coefficients for comparing the occurrence number with the fiber count (as in Figure 1), the fiber length with fiber count (as in Figure 2), and the occurrence number with fiber length (as in Figure 3). Then, we repeated the random label generations 5000 times.  From these 5000 random labeling experiments, the largest absolute values of the Spearman coefficients and the corresponding p-values are listed in Table 1.

\begin{table}
\begin{center}
\begin{tabular}{lcc}
 & Spearman coefficient & p-value \\ 
occurrence number - fiber count  & 0.0072 & 0.0232 \\
fiber length - fiber count & 0.0076 & 0.0167 \\ 
occurrence number - fiber length & 0.0062 & 0.0492
\end{tabular} 
\caption{The largest absolute values of the Spearman coefficients from the 5000 random labelings. In comparison with Figures 1, 2 and 3, the absolute values of the Spearman coefficients here is a little more than the 1\% of the values of the measured braingraph data. Consequently, the random controls do not have the relations between those parameters, demonstrated in Figures 1, 2, and 3.}
\end{center}
\end{table}
    
Clearly, even the largest absolute values of the Spearman coefficients of the 5000 random labellings are very small compared to those in the measured brain data.

\section*{Conclusions} 

By using our previously published robust braingraphs \cite{Varga2020} at \url{https://braingraph.org}, we have prepared a single consensus graph from 1064 graphs of 1015 vertices resolution, available at \url{https://braingraph.org/cms/download-pit-group-connectomes/}. Then, we compared the statistical similarity of the edge occurrence numbers, fiber lengths, and fiber counts of the edges. We have found a strong positive Spearman correlation between the edge occurrence numbers and the fiber count numbers, showing that statistically, the most frequent cerebral connections have the largest widths, i.e., the fiber number (Figure 1). We have found a negative Spearman correlation between the fiber lengths and fiber counts (Figure 2), showing that typically, the shortest edges are the widest or strongest by their fiber counts. We have also found a negative Spearman correlation between the occurrence numbers and the edge lengths in Figure 3: it shows that typically, the long edges are infrequent, and the frequent edges are short. By our knowledge, this is the first statistical comparison of these fundamental characteristics of the cerebral connections in more than 1000 braingraphs.
	
\section*{Data availability} All data are included in the text. 

\section*{Funding}

VG and DH were partially funded by the Ministry of Innovation and Technology of Hungary from the National Research, Development and Innovation Fund, financed under the  ELTE TKP 2021-NKTA-62 funding scheme.

\section*{Author Contribution} VG and DH have initiated the study and evaluated results; VG has overseen the work and wrote the first version of the paper; DH performed the data analysis and prepared the images, and both authors have reviewed the article.

\section*{Conflicting interest} The authors declare no conflicting interests.


\begin{thebibliography}{25}
\providecommand{\natexlab}[1]{#1}
\providecommand{\url}[1]{\texttt{#1}}
\expandafter\ifx\csname urlstyle\endcsname\relax
  \providecommand{\doi}[1]{doi: #1}\else
  \providecommand{\doi}{doi: \begingroup \urlstyle{rm}\Url}\fi

\bibitem[Szalkai and Grolmusz(2018)]{Szalkai2016b}
Bal{\'a}zs Szalkai and Vince Grolmusz.
\newblock Human sexual dimorphism of the relative cerebral area volumes in the
  data of the human connectome project.
\newblock \emph{European Journal of Anatomy}, 22\penalty0 (3):\penalty0
  221--225, 2018.

\bibitem[Hagmann et~al.(2008)Hagmann, Cammoun, Gigandet, Meuli, Honey, Wedeen,
  and Sporns]{Hagmann2008}
Patric Hagmann, Leila Cammoun, Xavier Gigandet, Reto Meuli, Christopher~J.
  Honey, Van~J. Wedeen, and Olaf Sporns.
\newblock Mapping the structural core of human cerebral cortex.
\newblock \emph{PLoS Biol}, 6\penalty0 (7):\penalty0 e159, Jul 2008.
\newblock \doi{10.1371/journal.pbio.0060159}.
\newblock URL \url{http://dx.doi.org/10.1371/journal.pbio.0060159}.

\bibitem[LaMontagne et~al.()LaMontagne, Benzinger, Morris, Keefe, Hornbeck,
  Xiong, Grant, Hassenstab, Moulder, Vlassenko, Raichle, Cruchaga, and
  Marcus]{LaMontagne2019}
Pamela~J. LaMontagne, Tammie~LS. Benzinger, John~C. Morris, Sarah Keefe, Russ
  Hornbeck, Chengjie Xiong, Elizabeth Grant, Jason Hassenstab, Krista Moulder,
  Andrei~G. Vlassenko, Marcus~E. Raichle, Carlos Cruchaga, and Daniel Marcus.
\newblock {OASIS}-3: Longitudinal neuroimaging, clinical, and cognitive dataset
  for normal aging and alzheimer disease.
\newblock \emph{medRxiv}.
\newblock \doi{10.1101/2019.12.13.19014902}.

\bibitem[McNab et~al.(2013)McNab, Edlow, Witzel, Huang, Bhat, Heberlein,
  Feiweier, Liu, Keil, Cohen-Adad, Tisdall, Folkerth, Kinney, and
  Wald]{McNab2013}
Jennifer~A. McNab, Brian~L. Edlow, Thomas Witzel, Susie~Y. Huang, Himanshu
  Bhat, Keith Heberlein, Thorsten Feiweier, Kecheng Liu, Boris Keil, Julien
  Cohen-Adad, M~Dylan Tisdall, Rebecca~D. Folkerth, Hannah~C. Kinney, and
  Lawrence~L. Wald.
\newblock The {H}uman {C}onnectome {P}roject and beyond: initial applications
  of 300 m{T}/m gradients.
\newblock \emph{Neuroimage}, 80:\penalty0 234--245, Oct 2013.
\newblock \doi{10.1016/j.neuroimage.2013.05.074}.
\newblock URL \url{http://dx.doi.org/10.1016/j.neuroimage.2013.05.074}.

\bibitem[Daducci et~al.(2012)Daducci, Gerhard, Griffa, Lemkaddem, Cammoun,
  Gigandet, Meuli, Hagmann, and Thiran]{Daducci2012}
Alessandro Daducci, Stephan Gerhard, Alessandra Griffa, Alia Lemkaddem, Leila
  Cammoun, Xavier Gigandet, Reto Meuli, Patric Hagmann, and Jean-Philippe
  Thiran.
\newblock The connectome mapper: an open-source processing pipeline to map
  connectomes with {MRI}.
\newblock \emph{PLoS One}, 7\penalty0 (12):\penalty0 e48121, 2012.
\newblock \doi{10.1371/journal.pone.0048121}.
\newblock URL \url{http://dx.doi.org/10.1371/journal.pone.0048121}.

\bibitem[Kerepesi et~al.(2017)Kerepesi, Szalkai, Varga, and
  Grolmusz]{Kerepesi2016b}
Csaba Kerepesi, Balazs Szalkai, Balint Varga, and Vince Grolmusz.
\newblock The braingraph. org database of high resolution structural
  connectomes and the brain graph tools.
\newblock \emph{Cognitive Neurodynamics}, 11\penalty0 (5):\penalty0 483--486,
  2017.

\bibitem[Varga and Grolmusz(2021)]{Varga2020}
Balint Varga and Vince Grolmusz.
\newblock The braingraph.org database with more than 1000 robust human
  structural connectomes in five resolutions.
\newblock \emph{Cognitive Neurodynamics}, 2021.
\newblock URL \url{https://doi.org/10.1007/s11571-021-09670-5}.

\bibitem[Keresztes et~al.()Keresztes, Szogi, Varga, and
  Grolmusz]{Keresztes2020}
Laszlo Keresztes, Evelin Szogi, Balint Varga, and Vince Grolmusz.
\newblock Introducing and applying newtonian blurring: An augmented dataset of
  126,000 human connectomes at braingraph.org.
\newblock \emph{Scientific Reports}.
\newblock \doi{10.1038/s41598-022-06697-4}.
\newblock URL \url{https://www.nature.com/articles/s41598-022-06697-4}.

\bibitem[Kerepesi et~al.(2018{\natexlab{a}})Kerepesi, Szalkai, Varga, and
  Grolmusz]{Kerepesi2015a}
Csaba Kerepesi, Bal{\'a}zs Szalkai, B{\'a}lint Varga, and Vince Grolmusz.
\newblock Comparative connectomics: Mapping the inter-individual variability of
  connections within the regions of the human brain.
\newblock \emph{Neuroscience Letters}, 662\penalty0 (1):\penalty0 17--21,
  2018{\natexlab{a}}.
\newblock \doi{10.1016/j.neulet.2017.10.003}.

\bibitem[Szalkai et~al.(2015{\natexlab{a}})Szalkai, Kerepesi, Varga, and
  Grolmusz]{Szalkai2015a}
Bal{\'a}zs Szalkai, Csaba Kerepesi, B{\'a}lint Varga, and Vince Grolmusz.
\newblock The {B}udapest {R}eference {C}onnectome {S}erver v2. 0.
\newblock \emph{Neuroscience Letters}, 595:\penalty0 60--62,
  2015{\natexlab{a}}.

\bibitem[Szalkai et~al.(2017{\natexlab{a}})Szalkai, Kerepesi, Varga, and
  Grolmusz]{Szalkai2016}
Balazs Szalkai, Csaba Kerepesi, Balint Varga, and Vince Grolmusz.
\newblock Parameterizable consensus connectomes from the {H}uman {C}onnectome
  {P}roject: The {B}udapest {R}eference {C}onnectome {S}erver v3.0.
\newblock \emph{Cognitive Neurodynamics}, 11\penalty0 (1):\penalty0 113--116,
  feb 2017{\natexlab{a}}.
\newblock \doi{http://dx.doi.org/10.1007/s11571-016-9407-z}.

\bibitem[Kerepesi et~al.(2018{\natexlab{b}})Kerepesi, Varga, Szalkai, and
  Grolmusz]{Kerepesi2016}
Csaba Kerepesi, Balint Varga, Balazs Szalkai, and Vince Grolmusz.
\newblock The dorsal striatum and the dynamics of the consensus connectomes in
  the frontal lobe of the human brain.
\newblock \emph{Neuroscience Letters}, 673:\penalty0 51--55, March
  2018{\natexlab{b}}.
\newblock \doi{10.1016/j.neulet.2018.02.052}.

\bibitem[Szalkai et~al.(2017{\natexlab{b}})Szalkai, Varga, and
  Grolmusz]{Szalkai2016e}
Bal{\'a}zs Szalkai, B{\'a}lint Varga, and Vince Grolmusz.
\newblock The robustness and the doubly-preferential attachment simulation of
  the consensus connectome dynamics of the human brain.
\newblock \emph{Scientific Reports}, 7\penalty0 (16118), 2017{\natexlab{b}}.
\newblock \doi{10.1038/s41598-017-16326-0}.

\bibitem[Szalkai et~al.(2019)Szalkai, Kerepesi, Varga, and
  Grolmusz]{Szalkai2016d}
Balazs Szalkai, Csaba Kerepesi, Balint Varga, and Vince Grolmusz.
\newblock High-resolution directed human connectomes and the consensus
  connectome dynamics.
\newblock \emph{PLoS ONE}, 14\penalty0 (4):\penalty0 e0215473, September 2019.
\newblock URL \url{https://doi.org/10.1371/journal.pone.0215473}.

\bibitem[Kerepesi et~al.(2016)Kerepesi, Szalkai, Varga, and
  Grolmusz]{Kerepesi2015b}
Csaba Kerepesi, Balazs Szalkai, Balint Varga, and Vince Grolmusz.
\newblock How to direct the edges of the connectomes: Dynamics of the consensus
  connectomes and the development of the connections in the human brain.
\newblock \emph{PLOS One}, 11\penalty0 (6):\penalty0 e0158680, June 2016.
\newblock URL \url{http://dx.doi.org/10.1371/journal.pone.0158680}.

\bibitem[Szalkai et~al.(2015{\natexlab{b}})Szalkai, Varga, and
  Grolmusz]{Szalkai2015}
Bal{\'{a}}zs Szalkai, B{\'{a}}lint Varga, and Vince Grolmusz.
\newblock Graph theoretical analysis reveals: Women's brains are better
  connected than men's.
\newblock \emph{PLoS One}, 10\penalty0 (7):\penalty0 e0130045,
  2015{\natexlab{b}}.
\newblock \doi{10.1371/journal.pone.0130045}.
\newblock URL \url{http://dx.doi.org/10.1371/journal.pone.0130045}.

\bibitem[Szalkai et~al.(2021)Szalkai, Varga, and Grolmusz]{Szalkai2016a}
Bal{\'a}zs Szalkai, B{\'a}lint Varga, and Vince Grolmusz.
\newblock The graph of our mind.
\newblock \emph{Brain Sciences}, 11\penalty0 (3), 2021.
\newblock URL \url{https://doi.org/10.3390/brainsci11030342}.

\bibitem[Szalkai et~al.(2018)Szalkai, Varga, and Grolmusz]{Szalkai2015c}
Bal{\'a}zs Szalkai, B{\'a}lint Varga, and Vince Grolmusz.
\newblock Brain size bias-compensated graph-theoretical parameters are also
  better in women's connectomes.
\newblock \emph{Brain Imaging and Behavior}, 12\penalty0 (3):\penalty0
  663--673, 2018.
\newblock \doi{10.1007/s11682-017-9720-0}.
\newblock URL \url{http://dx.doi.org/10.1007/s11682-017-9720-0}.

\bibitem[Fellner et~al.(2019)Fellner, Varga, and Grolmusz]{Fellner2017}
Mate Fellner, Balint Varga, and Vince Grolmusz.
\newblock The frequent subgraphs of the connectome of the human brain.
\newblock \emph{Cognitive Neurodynamics}, 13\penalty0 (5):\penalty0 453--460,
  2019.
\newblock URL \url{https://doi.org/10.1007/s11571-019-09535-y}.

\bibitem[Fellner et~al.(2020{\natexlab{a}})Fellner, Varga, and
  Grolmusz]{Fellner2019}
M{\'a}t{\'e} Fellner, B{\'a}lint Varga, and Vince Grolmusz.
\newblock The frequent complete subgraphs in the human connectome.
\newblock \emph{PloS One}, 15\penalty0 (8):\penalty0 e0236883,
  2020{\natexlab{a}}.
\newblock URL \url{https://doi.org/10.1371/journal.pone.0236883}.

\bibitem[Fellner et~al.(2020{\natexlab{b}})Fellner, Varga, and
  Grolmusz]{Fellner2018}
Mate Fellner, Balint Varga, and Vince Grolmusz.
\newblock The frequent network neighborhood mapping of the human hippocampus
  shows much more frequent neighbor sets in males than in females.
\newblock \emph{PLOS One}, 15\penalty0 (1):\penalty0 e0227910,
  2020{\natexlab{b}}.
\newblock URL \url{https://doi.org/10.1371/journal.pone.0227910}.

\bibitem[Fellner et~al.(2020{\natexlab{c}})Fellner, Varga, and
  Grolmusz]{Fellner2019a}
Mate Fellner, Balint Varga, and Vince Grolmusz.
\newblock Good neighbors, bad neighbors: The frequent network neighborhood
  mapping of the hippocampus enlightens several structural factors of the human
  intelligence on a 414-subject cohort.
\newblock \emph{Scientific Reports}, 10\penalty0 (11967), 2020{\natexlab{c}}.
\newblock URL \url{https://doi.org/10.1038/s41598-020-68914-2}.

\bibitem[Keresztes et~al.(2022)Keresztes, Szogi, Varga, and
  Grolmusz]{Keresztes2022a}
Laszlo Keresztes, Evelin Szogi, Balint Varga, and Vince Grolmusz.
\newblock Discovering sex and age implicator edges in the human connectome.
\newblock \emph{Neuroscience letters}, 791:\penalty0 136913, November 2022.
\newblock ISSN 1872-7972.
\newblock \doi{10.1016/j.neulet.2022.136913}.

\bibitem[Hegedus and Grolmusz(2023)]{Hegedus2022}
Daniel Hegedus and Vince Grolmusz.
\newblock Robust circuitry-based scores of structural importance of human brain
  areas.
\newblock \emph{PLOS One}, 2023.
\newblock \doi{https://doi.org/10.1371/journal.pone.0292613}.

\bibitem[Spearman()]{Spearman}
Carl Spearman.
\newblock The proof and measurement of association between two things.
\newblock \emph{The American Journal of Psychology}, 15\penalty0 (1):\penalty0
  72--101.
\newblock URL \url{https://doi.org/10.2307/1412159}.

\end{thebibliography}

\end{document}